\documentclass{article}
\begin{document}
\title{Analyticity and Quark-Gluon Structure of Hadrons}
\author{M. Majewski\thanks{Electronic address: mimajew@mvii.uni.lodz.pl}\\
University of Lodz, Department of Theoretical Physics\\
ul.~Pomorska 149/153, 90-236 Lodz, Poland\\
\and V. A. Meshcheryakov\thanks{Electronic address: mva@thsun1.jinr.ru}\\
Joint Institute for Nuclear Research\\
Dubna 141980, Moscow Region, Russia}
\maketitle

\begin{abstract}
The amplitudes of hadron-hadron
forward elastic scattering at high energy are investigated on the basis of
analiticity and crossing-symmetry which is valid in QCD. The universal
uniformizing  variable for them is proposed and the formulae for crossing-even 
and crossing-odd amplitudes are derived. The same parameters in these formulae
determine the real and imaginary (total cross sections) parts of the amplitudes.
 The analysis of the parameters determined from experimental data clearly
points to the quark-gluon stracture of hadrons. The total cross sections for
hyperon-proton scattering are predicted. They are consistent with
experimental data and, in particular, with the new SELEX-collaboration
measurement $\sigma_{tot}(\Sigma^{-} p)$.\\[\baselineskip]
\end{abstract}

\section{Introduction}

The modern theory of strong interactions (QCD) involves a number of unsolved
problems, for instance, the problem whether glueballs exist or not. Among
them, there is the problem of analytic properties of physical amplitudes. In  a
series of papers [1], Ohme has shown that for gauge theories quantized on
the basis of BRST algebra [2] the confinement conditions can be formulated
in such a way,  that physical amplitudes do possess the analytic properties and 
conditions
of crossing symmetry established earlier [3]. In particular, the forward $\pi 
p$ scattering amplitude has two nucleon poles and two cuts corresponding to 
direct and cross processes. The problem whether double dispersion
relations are valid or not remains still open in gauge theories.

Below, the Ohme's result is used to construct a model of the amplitude of
scattering of a hadron A on a proton. This model allows one to determine 
the quark-gluon structure of hadron A on the experimental basis, avoiding
controversial questions on the pomeron multicomponent structure [4-6]. 

\section{Universal Riemann surface of the forward scattering amplitude}

The notion of universal Riemann surface of forword scattering amplitude for
hadron-hadron processes at high energes arises when one introduces the well known 
variable
$$\nu=\frac{s-u}{4M \mu },$$ 
where s,u are usual Mandelstam varables and M,$\mu $ are the masses of colliding 
particles. Thresholds of any elastic 
hadron-hadron process corresponding to the direct and cross reactions in the s-plane
transform into the points $\nu=\pm1$. The thresholds of all inelastic processes
(direc and crossing) lie on the cuts $(-\infty,-1], [+1,+\infty)$. They make the Riemann surface
of scattering amplitude as function of $\nu$ infinitely-sheeted. This property 
of the Riemann surface  can be modelled by particular choice of the uniformizing
variable, the same for all hadron-hadron processes,
 \begin{equation}w(\nu)=1/\pi \cdot \arcsin(\nu) .\end{equation}
The Riemann surface  of function $w(\nu$) is just what we call the universal Riemann 
surface.
It has three branch points: two of square root type in the points $\nu=\pm1$ and of 
the
logarithmic type at infinity.  The function $w(\nu)$ is suitable for taking 
account of the crossing symmetry of amplitudes of hadron-hadron scattering. We 
choose the latter so that the equality 
 \begin{equation}\mbox I \mbox m \mbox F_{\pm}^{A}=\sigma_{tot}^{(\bar A p)}\pm
\sigma_{tot}^{(Ap)}.\end{equation}
be valid on the upper edge of the right-hand cut of $\nu$ plane; then, the 
condition of crossing symmetry is: 
  \begin{equation}\mbox F_{\pm}(\nu)=\pm  \mbox F_{\pm}(-\nu).\end{equation}
 Besides, the 
amplitudes obey the condition of reality 
 \begin{equation}\mbox F_{\pm}^{*}(\nu)=-\mbox F_{\pm}(\nu ^{*}).\end{equation}
In the $w$-plane, a physical sheet of the  $\nu$-plane is mapped into
the strip $|\mbox R\mbox e w|\leq1/2,$  whose boundaries are images 
of cuts of the $\nu$-plane. We call it the physical strip in 
the $w$-plane. Nonphysical sheets of the $\nu$-plane transform into strips 
$|\mbox R \mbox e(w\pm n)|\leq1/2, (n=1,2, \cdots).$  This clearly demonstrates that  
the universal Riemann surface is infinite-sheeted.  

Let $w=x+iy.$  Then, owing to eq. (3-4), on the boundary of the physical strip 
we find  \begin{equation}\mbox F_{\pm}^{*}(1/2+iy)=\mp \mbox F_{\pm}(-1/2+iy).\end{equation}
Let us expand the amplitudes $\mbox F_{\pm}(w)$ into Taylor series with 
the center at the point $w_{0}=iy_{0}$. It's convergence radius is determined 
by the distance from the point $w_{0}$ to the nearest pole corresponding to the 
resonance on an unphysical sheet. The same parameters of that expansion 
determine both the real and imaginary parts of amplitudes $\mbox F_{\pm}(w)$.  
Below, we will use only the imaginary parts of amplitudes (the total cross 
sections) that can be represented by the following converging power series
\begin{eqnarray} 
\mbox Im \mbox F_{+}(1/2+iy)=\sum_{n\geq 1}\left(\frac{1}{2}\right)^{2n-2}\sigma_{+}^{(n)}(y), 
\nonumber\\
\sigma_{+}^{(1)}(y)=\sum_{m\geq1}a_{m}(y-y_{0})^{m-1},\qquad \sigma_{+}^{(n)}(y)=\frac{(-1)^{n+1}}{(2n-2)!}\cdot \frac{d^{2n-2}\sigma_{+}^{(1)}(y)}{dy^{2n-2}},
\nonumber\\
\mbox Im \mbox F_{-}(1/2+iy)=\sum_{n\geq 1}\left(\frac{1}{2}\right)^{2n-1}\sigma_{-}^{(n)}(y),\\
\sigma_{-}^{(1)}(y)=\sum_{m\geq1}b_{m}(y-y_{0})^{m-1},\qquad \sigma_{-}^{(n)}(y)=\frac{(-1)^{n+1}}{(2n-2)!}\cdot\frac{d^{2n-2}\sigma_{-}^{(1)}(y)}{dy^{2n-2}}.
\nonumber
\end{eqnarray}
Expansions (6) satisfy equation (5).  It is instructive to compare the 
argument of expansions (6) with commonly used expressions, for instance: 
$(\frac{s}{s_{0}})^{\alpha}, s_{0}=1{\rm GeV}^2$ in refs.  [4,6,8] and $(p/20)^{\alpha}$ in ref. [9]
(here and everewere below p is the momentum in the lab. system). However, 
when one attempts to compare two different parametrizations of total cross 
sections in the region $\sim$ 100 {\rm GeV}/c, the function $\ln( p/p_{0})$ arises 
naturally.  Let us derive it from formula (6).  From (1) it follows that 
$y=\ln(\nu+\sqrt{\nu^{2}-1}).$  For $s\gg M^{2}$, we have $y\sim \ln(2p/\mu)$. 
In this case, the function $(y-y_{0})\sim 1/\pi\ln( p/p_{0})$ is the argument 
of expansions (6).  Here the quantity $p_{0}$ has clear mathematical 
meaning--- it is the center of the expansion into the Taylor series, and at the 
same time, physically, it makes p dimensionless.
We stress once more that formulae (6) are valid in the vicinity of point $y_{0}$,
and they cannot be used to estimate  behavior of cross sections when $s\to \infty$;
discussions on the pomeron structure refer to the region where they are not applicable.
\section{Quark-gluon structure of hadrons}

Formulae (6) were employed to analyze the experimental data on total cross
sections pp, $\bar pp,K^{\pm}p,\pi^{\pm}p.$ [7]. The results are collected in the
Table I. 
\begin{table}
\begin{center}
\caption{The values of the parameters $a_{m},b_{m}$(all in mb) and $y_{0}$ in eq. (6).}
\begin{tabular}{||c|c|c|c|c||} \hline ~~~~ & $pp$ & $\pi p$ 
& kp&np  \\ \hline $a_{1}$ & $ 84.51\pm 0.18$  &$49.77\pm 0.09$ &$41.03\pm 0.12$& $83.49\pm 0.36$ \\
\hline $a_{2}$ &$-4.85\pm 0.36$ & $1.92\pm 0.19$&$5.16\pm 0.25$ &$-3.48\pm 0.62$\\ 
\hline $a_{3}$ & $15.97\pm 0.7$ & $10.37\pm 0.34$ & $7.37\pm 0.48$ & $8.72\pm 1.48$\\ 
\hline $b_{1}$ & $8.52\pm 0.17$ & $1.62\pm0.07$ &$3.51\pm0.12$ & $7.85\pm0.26$\\ 
\hline 
$b_{2}$ & $-13.82\pm 0.79$ &$-2.8\pm 0.17$& $-5.65\pm 0.51$& $-12.74\pm 1.24$ \\ 
\hline $b_{3}$ & $15.33\pm 1.7$ &$2.7\pm 1.8$ &$5.04\pm 0.97$ & $12.36\pm 2.97$ \\ 
\hline $y_{0}$ & 1.71 & 2.31 & 1.91 & 1.71 \\ 
\hline $\chi_{n_{D}}^{2}$ & $ \frac{112}{109}$ & $\frac{82}{73}$ & 
$\frac{48}{38}$ & $\frac{96}{50}$ \\  
\hline \end{tabular}
\end{center}
\end{table}
Twenty four coefficients $a_{m},b_{m}$ are determined by 300 
experimental points and describe behaviour of cross sections in the 
interval $p\in(10,10^3)\,{\rm GeV}/c$. Values of $y_{0}$ correspond to $p=100\,
{\rm GeV}/c$, at which the correlations between parameters $a_{m},b_{m}$ are 
minimal.  In the vicinity of $y_{0}$, the considered total cross sections have
minima, and the real parts of amplitudes cross over the zero.  Twelve coefficients 
$b_{m}$ display the simple dependence:  $$(b_{m})_{pp}:(b_{m})_{\pi p}:(b_{m})_{Kp}:(b_{m})_{np}=5:1:2:4.$$ 
The mean ratios are calculated from Table 1
to be as follows:  $$\overline{\left(\frac{b_{pp}}{b_{\pi
p}}\right)}=5.37\pm0.22\qquad\overline{\left(\frac{b_{Kp}}{b_{\pi p}}\right)}=2.16\pm0.12
\qquad\overline{\left(\frac{b_{np}}{b_{\pi p}}\right)}=4.79\pm0.23.$$ 
They are in good agreement with ratios (7), except for the last one. It differs from
 (7) by three standard deviations as a result 
of large $\chi^{2}/n_{D}$ for np scattering.Therefore, it is expedient to 
use it below only for qualitative estimations. 
Relations (7) are not new and are written in order to demonstrate that the 
analysis of coefficients $a_{m},b_{m}$ is important for determining quark and 
other degrees of freedom of hadrons. It is known [9], that relationships (7) 
follow from consideration of annihilation components of amplitudes and are 
proportional to the numbers of dual diagrams of scattering of a hadron on 
proton  \begin{equation}n_{d}(Ap) =2N_{\bar u}^{A}+N_{\bar d}^{A} \end{equation}
where $N_{\bar u}^{A},N_{\bar d}^{A}$ are numbers of antiquarks $\bar u,\bar d$ in hadron 
A.

It is of great interest but difficult to analyze the crossing even part of 
the scattering amplitude.  The additive quark model (AQM) [10] predicts the 
following ratios $$\sigma_{pp}:\sigma_{\pi p}:\sigma_{Kp}:\sigma_{np}=3:2:2:3$$ 
However, from our Table 1 it is seen that only the 
coefficients $a_{1}$ and $a_{3}$ approximately follow that dependence.  
The difference $(a_{1})_{pp}-(a_{1})_{np}=1.02\pm0.40$ can be 
considered compatible with zero, since it does not exceed three standard deviations,
 and the description of process $np$ is not quite satisfactory. We will neglect the
 distinction between processes $pp$ and $np$, though, for the coefficient $a_{3}$, 
this assumption is valid only due to $\chi^2/n_{D}$ being large in magnitude.
 At the same time, the difference $(a_{1})_{\pi p}-(a_{1})_{kp}=8.74\pm0.15$ is significant 
and, together with other coefficients, determines 30 \% accuracy of the additive quark model. 
The values of coefficients $a_{2}$  from Table 1
do not comport with the AQM predictions, and therefore, they are very important 
for choosing new models. Some attempts of refining the AQM are known [11, 12].
 All of them suggest that the amplitude should be supplemented with terms bilinear in quark 
numbers of hadrons A. In this case, the amplitude can be described 
satisfactorily under different assumptions on the form of bilinear terms. 
However, their clear physical justification is rather difficult.

Below, we construct a new model by using the known idea of quarks being
confined in a hadron by gluons. Then it is natural to assume that the total
cross section of scattering of hadron A on a proton contains a part that
describes gluon-gluon interaction.
With this in mind, we set
 \begin{equation}a_{m}=\alpha_{m}+\beta_{m} \cdot N^{A}_{q}+\gamma_{m} \cdot N^{A}_{q} \cdot N^{A}_{ns} \end{equation}
where $N^{A}_{q}$ is the total number of quarks; $N^{A}_{ns}$ is the total
 number of nonstrange quarks in hadron A; and the numbers $\alpha_{m}$ do not 
depend on the quark content of hadron A [9]. The numbers $\alpha_{m}$ determine the fraction of the total cross section
corresponding to the gluon-gluon interaction.  It is just the gluon degree of
freedom of hadrons A and p that is responsible for them. The assumption on $a_m$
(eq.(8)) corresponds to the hypothesis of Gershtein and Logunov [13]. They argue that 
the constant of Froissart limit doesn't depend on the guark content of hadron A ,
but it does depend on glueballs and is the same for all processes. The hypothesis 
has been verified by Prokoshkin [14] on the basis of similar experimetal data as we use.
In our model one should attribute the Froissart behavior not to the variable y but to 
$y_0$ one. That justifies the eq.(8).
 The different numbers $(a_{m})_{pp},(a_{m})_{\pi p},(a_{m})_{kp}$ determine $\alpha_{m},\beta_{m},\gamma_{m}.$ 
Then, the prediction power of hypothesis (8) can be verified for the values of total cross sections of
hyperon-proton interactions. In ref. [15], the results are presented on the
measurement of total cross sections of $\Sigma^{-}p$ and $\Xi^{-}p$ in the range of momenta
$(74.5, 136.9)\,{\rm GeV}/c$. In this range, the total cross sections are varying slightly, and to
compare the predictions of the model given by formulae (1), (6), and (8), we
take the momentum $p=101\,{\rm GeV}/c$. In this case, the theoretical and experimental results
are as follows:
$$
\sigma_{\Xi^- p} = \begin{array}{l}(29.25 \pm 0.5\, {\rm mb})_{th}\\ 
(29.12 \pm 0.22\, {\rm mb})_{ex}\end{array}, \quad
\sigma_{\Sigma^- p} = \begin{array}{l}( 34.8 \pm 0.2\, {\rm mb})_{th}\\ 
(33.3 \pm 0.3\, {\rm mb})_{ex}\end{array}
$$
Similar data [16] for $\Lambda p$ and $\Sigma^{-} p$ scattering at $20\,{\rm GeV}/c$ are
$$
\sigma_{\Lambda p} = \begin{array}{l} (33.3 \pm 0.5\, {\rm mb})_{th}\\ 
(34.7 \pm 3\, {\rm mb})_{ex}\end{array}, \quad
\sigma_{\Sigma^- p} = \begin{array}{l} (34.2 \pm 0.5\, {\rm mb})_{th}\\ 
(34 \pm 1\, {\rm mb})_{ex}\end{array}
$$
Recently, the collaboration SELEX has published the data on $\Sigma^- p$ at $p=609\,
{\rm GeV}/c.$ [17].
The comparison with predictions of the model is
$$
\sigma_{\Sigma^- p}=\begin{array}{l} (35 \pm 7.5\, {\rm mb})_{th}\\
(37 \pm 0.7\, {\rm mb})_{ex}\end{array}
$$
Though the obtained value of the total cross section is not so accurate as in
ref. [18], it should be considered satisfactory. In refs. [4, 5, 6, 11, 18]
devoted to the analysis of total cross sections, the errors of predicted
values were not calculated, but they increase rapidly in the region of
extrapolation.
\section{Conclusion}

A uniformizing variable for hadron-hadron forward scattering at high energes 
was proposed on the basis of analyzing the analytic properties of physical scattering
amplitudes [1].  If one represents the scattering amplitudes as Taylor series
in that variable and takes crossing symmetry into account, one can once more
be convinced on the experimental basis that hadrons possess the quark-gluon
structure.  Predicted the total cross sections for scattering of strange hadrons
on proton are in agreement with experiment in a wide energy range. The
gluon-gluon part of the total cross sections at momenta $p=100\,{\rm GeV}/c$ amounts
to about 10\%.


\begin{thebibliography}{99}
\bibitem{1} R. Oehme, Int. J. of Mod. Phys. {\bf A10}, 1995 (1995) and references therein.
\bibitem{2} C. Becchi, A. Rouet, R. Stora, Ann. Phys. {\bf 98}, 2879 (1976);\\
 I.V. Tyutin, Lebedev Report , FIAN 39 (1975).
\bibitem{3} N.N. Bogoliubov, B.V. Medvedev, M.K. Polivanov,
\\Voprosy Teorii Dispersionnykh Sootnoshenii, Fizmatgiz, Moscow, 1958.
\bibitem{4} A. Donnachie, P.V. Landshoff, Phys. Lett. {\bf B296}, 227 (1992).
\bibitem{5} H.J. Lipkin, Phys. Lett. {\bf B335}, 500 (1994).
\bibitem{6} P. Gauron, B. Nicolescu, A possible two-component structure \\
of the non-perturbative Pomeron, preprint LPNE 00-02, Paris.
\bibitem{7} V.P. Gerdt, V.I. Inozemtsev, V.A. Meshcheryakov, \\Lett. Nuovo
Cimento {\bf 15}, 321 (1976).
\bibitem{8} Particl Data Grup, Europian Physics Jornal {\bf C3}, 205 (1998).
\bibitem{9} H.J. Lipkin, Phys. Rev. {\bf  D11}, 827 (1975].
\bibitem{10} E.M. Levin, L.L. Frankfurt, Pisma JETP {bf 2}, 105 (1965).
\bibitem{11} P. Jonson, B. Nicolescu, Nuovo Cim. {\bf  A37}, 97 (1977).
\bibitem{12} H.J. Lipkin, Nucl. Phys. {\bf  B78}, 1381 (1974).
\bibitem{13} S.S. Gershtein, A.A. Logunov, Yad. Fiz. {\bf 39}, 1514 (1984).
\bibitem{14} Y.D. Prokoshkin Yad. Fiz. {\bf 40}, 1579 (1984).
\bibitem{15} S. Gjesdal et al., Phys. Lett. {\bf B40}, 152 (1972).
\bibitem{16} S.F. Biagis et al., Nucl. Phys. {\bf  B186}, 1 (1981).
\bibitem{17} U.Dersch et al., Nucl. Phys. {\bf B579}, 277 (2000).
\bibitem{18} H.J. Lipkin, The New $\sigma_{tot}(\Sigma p)$ data, the new PDG fit
to hadron total cross section  and the TCP alternative, hep-ph/9911259.
\end{thebibliography}
\end{document}